\newcommand\Ra{\mbox{\textit{Ra}}}  % Reynolds number
\newcommand\Sc{\mbox{\textit{Sc}}}  % Reynolds number

\newcommand{\parone}[2] {\df{\partial #1}{\partial#2}}

\newcommand{\bld}[1]{\boldsymbol{#1}}

\newcommand{\df}[2]{\displaystyle{\frac{#1}{#2}}}

\documentclass{jfm}
\usepackage{subcaption}
\usepackage[font=footnotesize]{subcaption} 
\usepackage{amssymb}
\usepackage{multirow}
\usepackage{array}
\usepackage{graphicx}
\usepackage{epstopdf, epsfig}
\usepackage{color}
\usepackage{amsmath,mathtools}
\usepackage{placeins}
\usepackage{changes}
\allowdisplaybreaks

\shorttitle{The instability of gyrotactically-trapped cell layers}
\shortauthor{S. Maretvadakethope, E. E. Keaveny and Y. Hwang}

\title{The instability of gyrotactically-trapped cell layers}

\author{Smitha Maretvadakethope\aff{1},
   Eric E. Keaveny\aff{1}
 \and Yongyun Hwang\aff{2}\corresp{\email{y.hwang@imperial.ac.uk}}}

\affiliation{\aff{1}Department of Mathematics, Imperial College London,\\
South Kensington, London SW7 2AZ, UK
\aff{2}Department of Aeronautics, Imperial College London, \\
South Kensington, London SW7 2AZ, UK}

\begin{document}

\maketitle

\begin{abstract}
Several meters below the coastal ocean surface there are areas of high ecological activity that contain thin layers of concentrated motile phytoplankton. Gyrotactic trapping has been proposed as a potential mechanism for layer formation of bottom-heavy swimming algae cells, especially in flows where the vorticity varies linearly with depth (Durham, Stocker \& Kessler, \emph{Science}, vol. 323, 2009, pp. 1067-1070). Using a continuum model for dilute microswimmer suspensions, we report that an instability of a gyrotactically trapped cell-layer can arise in a pressure-driven plane channel flow.  The linear stability analysis reveals that the equilibrium cell-layer solution is hydrodynamically unstable due to negative microswimmer buoyancy (i.e. a gravitational instability) over a range of biologically relevant parameter values. {The critical cell concentration for this instability is found to be $N_c\simeq 10^4\textrm{cells}/\textrm{cm}^{3}$, a value comparable to the typical maximum cell concentration observed in  thin layers. This result indicates that the instability may be a potential mechanism for limiting the layer's maximum cell concentration, especially in  regions where turbulence is weak, and motivates the study of its  nonlinear evolution, perhaps, in the presence of turbulence.} 
\end{abstract}

\begin{keywords}
%Authors should not enter keywords on the manuscript, as these must be chosen by the author during the online submission process and will then be added during the typesetting process (see http://journals.cambridge.org/data/\linebreak[3]relatedlink/jfm-\linebreak[3]keywords.pdf for the full list)
\end{keywords}
\section{Introduction}

Oceanographic studies have shown that near coastal regions long and thin layers of phytoplankton form several metres beneath the surface and persist for days {\citep{nielsen1990effects,  dekshenieks2001temporal, stacey2007convergences, sullivan2010coastal}. These layers have cell concentrations orders of magnitude higher than ambient values \citep{sullivan2010coastal} and are ecological hot spots that significantly contribute to species diversity in marine environments \citep{grunbaum2009peter}. Their thickness ranges from a few centimetres to a few meters and they can extend horizontally for kilometres \citep{dekshenieks2001temporal, moline2010integrated}.}
% Multiple patches have been shown to coexist simultaneously at various depths \citep{rines2010thin} allowing for increased marine diversity, and opportunities for different species to specialise \citep{grunbaum2009peter}. Depending on whether the phytoplankton are toxic or not, layer formation has also been seen to increase algal blooms and zooplankton and fish mortality rates \citep{nielsen1990effects, townsend2005nature, bjornsen1991decimeter, donaghay1997toward, yamochi1984mechanisms}, or increase prey concentrations required for fish larvae to survive \citep{lasker1975field}.} %Additionally, layers may affect optical and acoustic properties of the ocean \citep{dekshenieks2001temporal,alldredge2002occurrence}.}%mcmanus2003characteristics,
%While it is known that turbulence can destabilise these layers, it has not been studied how layer stability relates to swimmer motility or buoyancy \citep{stacey2007convergences}.
{These layers often emerge in  areas where  turbulence is considerably weak or highly suppressed by surrounding flow conditions (e.g. density stratification) \citep{dekshenieks2001temporal}. A number of mechanisms have been proposed for layer formation, and they include convergent swimming originating from `taxes' of motile phytoplankton \citep{macintyre1997vertical,grunbaum2009peter,ryan2010interacting}, vertical stratification in ocean especially at pycnoclines \citep{dekshenieks2001temporal,johnston2009observations}, straining by shear \citep{osborn1998finestructure,  franks1995thin, stacey2007convergences} and gyrotactic trapping \citep{durham2009disruption}. The reader may refer to \cite{durham2012thin} for a review on this issue.} 

%Convergent swimming is the locomotion of various swimmers towards a stimulus  \citep{durham2012thin}. Examples of swimming targets include areas with nutrients \citep{ryan2010interacting} or preferred salinities \citep{grunbaum2009peter}, or even areas with a preferred level of light \citep{macintyre1997vertical}. Layers are also formed due to vertical stratification as sinking cells can become neutrally buoyant at pycnoclines \citep{dekshenieks2001temporal}, where vertical shear has been shown to be largest \citep{johnston2009observations}. Straining cell populations by the shear is also known to be responsible for layer formation \citep{osborn1998finestructure,  franks1995thin, stacey2007convergences}.} %eckart1948analysis,

{Of particular interest to the present study is the mechanism of gyrotactic trapping, which has been proposed for green algae, like those of genus \emph{Chlamydomonas} or \emph{Dunaliella}. \citep*{durham2009disruption}. These microorganisms are commonly observed in  lakes, seas and oceans \citep{ginzburg1985influence, krivtsov2000changes, simila1988spring} and they exhibit sensitivity to gravity \citep{kessler1984gyrotactic}.} This is the consequence of bottom heaviness originating from the fact that the centre of mass of the cell is displaced from that of buoyancy.  This displacement results in a gravitational torque causing the cell to swim upwards \citep{kessler1985hydrodynamic}. In the presence of shear, however, such a cell also experiences a viscous torque due to the vorticity of the flow, and the swimming direction is then determined by the balance between the gravitational and viscous torques (i.e. gyrotaxis). In particular, if the shear %and the corresponding vorticity are 
is very large, the cell continuously tumbles \citep{pedley1987orientation} and {gradually loses its upswimming velocity on average.} 

{\cite{durham2009disruption} proposed that an excessively large shear (or vorticity) disrupts the upswimming of large numbers of gyrotactic cells, thereby leading to the formation of a thin layer in the region of large shear (i.e. gyrotactic trapping). To demonstrate this mechanism, they performed a laboratory experiment where a flow with an approximately linearly growing shear rate is applied to a suspension of gyrotactic cells (\emph{C. nivalis} and \emph{H. akashiwo}) with $1cm$ depth. %, using a rotating Mylar belt located at the top.
 It was shown that a thin layer of cells is indeed formed by the proposed mechanism. However, 
the layer was also found} to be highly unsteady and exhibit non-trivial dynamics {in the sense that the formation of the layer is highly transient \cite[see also numerical study by][]{santamaria2014gyrotactic}}, even though the background flow itself remained laminar and steady (private communication with W. M. Durham). 
{In bioconvection, a thin layer is  formed  at the upper  fluid boundary due to the upswimming of the cells. This layer has been found to become unstable as the high cell concentration causes gravitational overturning, which in turn causes a convection pattern to arise \cite[e.g.][]{pedley1988growth,pedley1992hydrodynamic,bees1997wavelengths}. Similarly, the thin layers formed in the suspension by gyrotactic trapping are suspected to exhibit a similar instability, which may explain the highly unsteady layer dynamics observed in experiments.} %Comparable trapping due to fluid shear has been observed for bacterial flow suspensions, where transport was suppressed due to cell depletion in low shear regions \citep{rusconi2014bacterial}.}
%It is well known that in bioconvection, such thin layers exhibit instability as a consequence of gravitational overturning due to the layer being more dense than the ambient fluid \cite[for the more precise mechanism, see][]{ref2hwang2014bioconvection}. The thin layer formed by gyrotactic trapping is therefore expected to exhibit a similar hydrodynamic instability, which may explain the highly unsteady layer dynamics observed in experiments. 

The objective of the present study is to examine the stability of layers formed by gyrotactic trapping using the continuum model described in \cite{pedley2010instability} and \cite{ref2hwang2014bioconvection}. To this end, we consider a suspension of gyrotactic microorganisms subject to a horizontal plane Poiseuille flow which has a shear rate (or vorticity) that grows linearly in the vertical direction, as in the experiments by \cite{durham2009disruption}. {To some extent, this flow configuration appears to be similar to that of the experiment by \cite{ref1croze2010sheared} in which a horizontal pipe Poiseulle flow was  applied  to a suspension of  \emph{C. augustae}. However, unlike the pipe flow, the shear in horizontal plane Poiseuille flow is purely vertical, and we will see that this flow geometry admits a steady equilibrium solution in the form of a gyrotactically-trapped cell layer. This is the important physical feature which distinguishes the present study from the work by \cite{ref1croze2010sheared} as well as the one by \cite{ref2hwang2014bioconvection} which studied the role of uniform shear in bioconvection: the pressure driven channel flow has an equilibrium solution corresponding to a gyrotactically-trapped cell layer whose stability can then be assessed. Finally, it is worth mentioning that the layer formation in the plane Poiseulle flow of gyrotactic microorganism suspensions is reminiscent of that found for bacterial suspensions in a microfluidic channel \citep{rusconi2014bacterial}. However, the underlying mechanism for the layer formation between the two cases is fundamentally different.}  

\section{Problem formulation}\label{Problem formulation}
\subsection{Equations of motion}
The mathematical model and its stability analysis in the present study are based on those described in \citet{ref2hwang2014bioconvection}. {This model is derived from the Navier-Stokes equations combined with a Smolouchowski equation describing the cell distribution in time, space and cell-orientation space. The model adopts the translational diffusivity expression proposed by \cite{pedley1990new}, although this can be improved by the generalised Taylor's disperison theory \cite[e.g.][]{hill2002taylor}.} In our presentation here, we have omitted repeated details for brevity. Suppose we have a fluid of density, $\rho$, and kinematic viscosity, $\nu$, in an infinitely long and infinitely wide channel, subject to a constant pressure gradient in the horizontal direction. Here, we denote $x^*$, $y^*$ and $z^*$ as the streamwise, vertical and spanwise directions, respectively, and $t^*$ as the time (note that the superscript $^*$ denotes dimensional variables). The two walls of the channel are located at $y^*=\pm h$ where $h$ is half-height of the channel. In this horizontal channel there is a suspension of spherical gyrotactic cells with average cell number density, $N$. The individual cell is assumed to swim at speed $V^*_c$ and is subject to gravity, as well as diffusion.  The cell sedimentation speed is given by $V^*_s$, and, as we shall see in \S \ref{sec:basicstate}, plays a crucial role in the formation of the gyrotactically trapped equilibrium layer. The swimming direction of the cells is denoted by a unit vector, $\bld{e}=(e_1,e_2,e_3)$, and their sedimentation direction is $-\boldsymbol{j}$ (where $\boldsymbol{j}$ is the upward unit vector in the vertical direction). As shown in \cite{pedley1990new}, the translational diffusivity of the suspension scales like ${V_c^*}^2 \tau$, where $\tau$ is the swimming direction correlation time.  We use this as the representative value for translational diffusivity, such that $D_V\equiv {V_c^*}^2 \tau$. Using the length scale $h$, the diffusion velocity scale $D_V/h$, and the average cell number density $N$, the dimensionless equations of motion are: %\cite[for further details, see][]{ref2hwang2014bioconvection}: %HERE, ALL THE BOLD NABLA NEED TO BE CHANGED TO THE NORMAL ONE
\begin{subequations}
\label{nondimeqns}
\begin{gather}
\nabla\cdot \boldsymbol{u}=0,\label{nondimcont}
\\
\Sc^{-1}\left(\parone{\boldsymbol{u}}{t}+(\boldsymbol{u}\cdot \nabla) \boldsymbol{u}\right)=-\nabla p+\nabla^2 \boldsymbol{u}-\Ra~n\boldsymbol{j}\label{nondimmoment},
\\
\parone{n}{t}+\nabla\boldsymbol{\cdot}[n(\boldsymbol{u}+V_c\langle\boldsymbol{e}\rangle-V_s\boldsymbol{j})]=\nabla\boldsymbol{\cdot}(\mathsfbi{D}_T\boldsymbol{\cdot}\nabla n),\label{nondimadvec}
\\
\intertext{with the no-slip and no-flux boundary conditions on the walls}
\bld{u}|_{y=\pm 1}=(0,0,0), \label{nondimBC1}
\\
[n(\bld{u}+V_c\langle\bld{e}\rangle-V_s\bld{j})-\mathsfbi{D}_T\boldsymbol{\cdot }\nabla n]\large|_{y=\pm 1}\cdot \bld{j}=0, \label{nondimBC2s}
\end{gather}
where $\bld{u}(=(u,v,w))$, $p$, $n$, $V_c$, $V_s$ and $\mathsfbi{D}_T(\equiv \langle \boldsymbol{ee} \rangle - \langle \boldsymbol{e}\rangle\langle \boldsymbol{e}\rangle$; see also \cite{pedley1990new} for this expression), are the dimensionless velocity, pressure, cell number density, cell swimming speed, cell sedimentation speed, and translational diffusivity tensor, respectively. The Schmidt number $Sc$ and the Rayleigh number $Ra$ in (\ref{nondimmoment}) are given by
\begin{align}
&&\Sc=\dfrac{\nu}{D_V}~~\mathrm{and}~~\Ra=\dfrac{N\upsilon g'h^3}{D_V\nu},&&
\end{align}
where $g'={g\Delta \rho}/{\rho}$ is the reduced gravity ($\Delta \rho$ is the density difference between the cell and fluid), and $\upsilon$ the volume of a cell.
In (\ref{nondimadvec}) and (\ref{nondimBC2s}), $\langle \cdot \rangle$ denotes the local ensemble average at given spatial location $\boldsymbol{x}$, which is obtained with the p.d.f. of the cell swimming orientation, $f(\boldsymbol{x},\boldsymbol{e},t)$. This satisfies
\begin{equation}
D_R^{-1}\parone{f}{t}+D_R^{-1}(\bld{u\cdot }\nabla)f +\nabla_e\boldsymbol{\cdot }\left(\lambda(\bld{j}-(\bld{j}\cdot \bld{e})\bld{e})f+\dfrac{\boldsymbol{\Omega}}{2D_R}\times \bld{e}f\right)
=\nabla_e^2f, \label{nondimfeq} 
\end{equation}
where $\boldsymbol{\Omega}$ is  the flow vorticity. Here, the dimensionless rotational diffusivity $D_R$ and the dimensionless inverse of the gyrotactic time scale $\lambda$ are also given by
\begin{align}
&&D_R \equiv \frac{D_R^* h^2}{D_V}~~\mathrm{and}~~\lambda=\frac{1}{2BD_R^*},&&
\end{align}
\end{subequations}
with the rotational diffusivity $D_R^*$ and the gyrotactic time scale $B$.

\subsection{Basic State}
Given the horizontal homogeneity of the flow, (\ref{nondimeqns}) admits the following equilibrium solution:
\begin{subequations}
\begin{align}
\bld{u}_0(y)=(U_0(y),0,0),~~~n=n_0(y), \label{baseflowsimplif}
\end{align}
where
\begin{gather}
U_0(y)=\Sc\Rey(1-y^2), \label{basefloweqn}
\\
n_0(y)=N_0\exp \left(\int \dfrac{V_c\langle e_2\rangle_0-V_s}{D_{T0}^{22}}\mathrm{d}y \right). \label{basefloweqn2}
\end{gather}
Here, $Re=U_ch/\nu$ is the Reynolds number based on the centreline velocity $U_c$, and $N_0$ is the normalisation constant setting the volume average of $n_0(y)$ to be unity. Also, in (\ref{basefloweqn2}), $\langle \cdot \rangle_0$ is obtained from the steady and horizontally uniform solution $f_0$ of (\ref{nondimfeq}): i.e. 
\begin{align}\label{Baseq}
&\nabla_e\boldsymbol{\cdot} \left(\lambda(\bld{j}-(\bld{j}\cdot \bld{e})\bld{e})f_0+\dfrac{\boldsymbol{\Omega}_0}{2D_R}\times \bld{e}f_0\right)
=\nabla_e^2f_0, %\label{baseqpdf}
\end{align}
\end{subequations}
where $\boldsymbol{\Omega}_0=(0,0,-ScRe(\mathrm{d}U_0/\mathrm{d}y))$.

{In the regime of high shear rates, one may estimate the formation time scale of the basic state from an initially uniform suspension. In this case, the effect of the up-swimming velocity would be negligible due to very high surrounding shear. Therefore, the time scale for layer formation will be given by the length scale of the system and translational diffusivity, such that
\begin{equation}
T_{layer}\sim \frac{h^2}{D_V^*}=\frac{(0.25)^2}{1.98\times 10^{-4}}\approx 300s. \nonumber
\end{equation}}

\subsection{Linear stability analysis}
Now, let us consider a small perturbation around the basic state, such that: $[~\boldsymbol{u}~n~f~]^T= [~\boldsymbol{u}_0~n_0~ f_0~]^T+\epsilon[~\boldsymbol{u}'~ n'~f'~]^T+O(\epsilon^2)$ for $\epsilon \ll 1$. The normal-mode solution of the perturbation is then written as 
\begin{align}
[~\boldsymbol{u}'~ n'~f'~]^T(x,y,z,t)&=[~\hat{\boldsymbol{u}}~ \hat{n}~\hat{f}~]^T\mathrm{e}^{i(\alpha x+\beta z-\omega t)}+c.c,
%\eta'(x,y,z,t)&=\hat{\eta}(y) \mathrm{e}^{i(\alpha x+\beta z-\omega t)}+c.c.\\
%n'(x,y,z,t)&=\hat{n}(y) \mathrm{e}^{i(\alpha x+\beta z-\omega t)}+c.c.
\end{align}
where $\omega$ is the unknown complex angular frequency, $\alpha$ and $\beta$ are the given streamwise and spanwise wavenumbers, respectively. Using the standard procedure that eliminates pressure perturbation, the vertical velocity and vorticity form of the linearised equations of motion are given by 
\begin{subequations}\label{lns}
\begin{gather}
i\omega \begin{pmatrix}
\Sc^{-1}(k^2-\mathcal{D}^2)& 0 &0\\
0& \Sc^{-1}&0\\
0&0&1
\end{pmatrix}
\begin{pmatrix}
\hat{v}\\
\hat{\eta}\\
\hat{n}
\end{pmatrix}=
\begin{pmatrix}
L_{OS}& 0 &k^2 \Ra\\
i\beta \Rey\mathcal{D}U& L_{SQ}&0\\
\mathcal{D}n_0+L_C^v& L_C^\eta& L_C
\end{pmatrix}
\begin{pmatrix}
\hat{v}\\
\hat{\eta}\\
\hat{n}
\end{pmatrix},\label{matrixeqns}
\end{gather}
where 
\begin{gather}
L_{OS}= \mathrm{i}\alpha \Rey U(k^2-\mathcal{D}^2)+\mathrm{i}\alpha \Rey \mathcal{D}^2 U+(k^2-\mathcal{D}^2)^2,\\
L_{SQ}=\mathrm{i}\alpha \Rey  U+(k^2-\mathcal{D}^2),
\\
L_c=\mathrm{i}\alpha \Sc\Rey  U+\mathrm{i}\alpha V_c\langle e_1\rangle_0+(V_c\langle e_2\rangle_0-V_s)\mathcal{D}+\mathrm{i}\beta V_c\langle e_3\rangle_0+\text{$V_c\mathcal{D}\langle e_2\rangle_0$}\nonumber
\\
+\alpha^2D_{T0}^{11}-2\mathrm{i}\alpha D_{T0}^{12}\mathcal{D}-D_{T0}^{22}\mathcal{D}^2+\beta^2D_{T0}^{33} \text{$-\mathrm{i}\alpha\mathcal{D}{D_{T0}^{12}}-\mathcal{D}{D_{T0}^{22}}\mathcal{D}$},
\\
L_C^v =\Big[ G_1\mathcal{D}{n_0}\dfrac{\xi_{2}}{k^2}\mathrm{i}\alpha +n_0G_1\big(-\alpha^2\dfrac{\xi_{1}}{k^2} +\mathcal{D}\xi_{2}\dfrac{\mathrm{i}\alpha}{k^2} +\dfrac{\xi_{2}}{k^2} \mathcal{D}\mathrm{i}\alpha+\beta^2\dfrac{\xi_{3}}{k^2}\big)~~~~~~~~~~~~~~~~~~ \nonumber\\
~~~~-G_2\Big(\mathrm{i}\alpha\dfrac{\xi_{6}}{k^2} \mathcal{D}^2 n_0+\mathcal{D}n_0\big(-\alpha^2\dfrac{\xi_{5}}{k^2} +\mathcal{D}\xi_{6}{y}\dfrac{\mathrm{i}\alpha}{k^2} + \dfrac{\xi_{6}}{k^2} \mathcal{D}\mathrm{i}\alpha +\beta^2\dfrac{\xi_{7}}{k^2}\big)
\Big)\Big](k^2-\mathcal{D}^2),\\
L_C^\eta=G_1\mathcal{D}{n_0}\dfrac{\xi_{2}}{k^2}\mathrm{i}\beta \mathcal{D}+n_0G_1\Big[\big(\mathcal{D}\xi_{2}\dfrac{\mathrm{i}\beta}{k^2}+\dfrac{\xi_{2}}{k^2}\mathcal{D}\mathrm{i}\beta -\alpha\beta\dfrac{\xi_{3}}{k^2}
 -\alpha\beta\dfrac{\xi_{1}}{k^2}   \big)\mathcal{D}+\xi_{4}\mathrm{i}\beta\Big]~~~~~~~~~\nonumber\\
 ~~~-G_2\Big[\mathcal{D}n_0\big(\xi_{8}\mathrm{i}\beta+\Big(-\alpha\beta \dfrac{\xi_{5}}{k^2} + \dfrac{\mathrm{i}\beta}{k^2}\mathcal{D}\xi_{6}
+\dfrac{\xi_{6}}{k^2}\mathcal{D}\mathrm{i}\beta- \alpha\beta\dfrac{\xi_{7}}{k^2}\Big)\mathcal{D} \big)+
\mathrm{i}\beta\dfrac{\xi_{6}}{k^2}\mathcal{D}^2 n_0 \mathcal{D}\Big],
\end{gather}
with the boundary conditions %{\color{red} YH: EQUATIONS (2.4F) IS STRANGE. PLEASE CHECK IT ONCE AGAIN}
\begin{gather}
\hat{v}|_{y=\pm 1}=\mathcal{D}\hat{v}|_{y=\pm 1}=\hat{\eta}|_{y=\pm 1}=0,
\\
\bigg[
\big(V_c\langle{e}_2\rangle_0-V_s-i\alpha{D}_{T0}^{12} \big)\hat{n}-{D}_{T0}^{22}\mathcal{D}\hat{n}
+(G_1\xi_{2}n_0
-G_2\xi_{6}\mathcal{D}{n_0}
)\nonumber\\
\times
\big(\dfrac{i\alpha}{k^2} (k^2 
-\mathcal{D}^2)\hat{v}+ \dfrac{i\beta}{k^2} \mathcal{D}\hat{\eta}\big)
\bigg]_{y=\pm 1}=0.
\end{gather}
\end{subequations}
Here, $\mathcal{D}=\mathrm{d}/\mathrm{d}{y}$, $k^2=\alpha^2+\beta^2$, $\hat{\eta}=i\beta \hat{u}-i\alpha\hat{w}$, and $\xi_i$ are coefficients obtained in \cite{ref2hwang2014bioconvection} 
by applying a quasi-steady and quasi-uniform approximation to the linearised equation for $f'$. {We note that this approximation would be valid as long as the instability does not carry a flow perturbation with very small time and length scales \citep{ref2hwang2014bioconvection}. Fortunately, in laminar flows, such a perturbation would be damped by viscosity and diffusivity.} In (\ref{lns}), $G_1=V_c/D_R$ and $G_2=1/D_R$, and they indicate the relative importance of the cell swimming velocity and translational diffusion to rotational diffusion, respectively. %For further details on the derivation of (\ref{lns}), the reader may refer to \cite{ref2hwang2014bioconvection}.

The eigenvalue problem (\ref{lns}) is solved numerically by modifying the numerical solver used in \cite{ref2hwang2014bioconvection}. Here, the derivatives in the vertical direction are discretised using a Chebyshev collocation method and in the present study, all computations were carried out with 101 collocation points (i.e. $N_y = 101$).  We also note that no discernible changes were found for several test cases when increasing to $N_y=201$.

\subsection{Model parameters}
\begin{table}
  \begin{center}
\def~{\hphantom{0}}
  \begin{tabular}{lcc}
   Parameter    & Description   &   Reference Value              \\[3pt]
   $~Sc~$     & Schmidt number &   $~50.39~$  \\
   $~Ra~$     & Rayleigh number &   $~10^{0} -10^{4}~$  \\
   $~G_1~$        & see text &   $~0.38~$  \\
   $~G_2(=1/D_R)~$        & see text &   $~0.05~$  \\
   $~\lambda~$     & Inverse of dimensionless gyrotactic time scale &   $~2.2~$  \\
     $~Re~$        & Reynolds number of base-flow shear &  {$~0 - 6.28~$ } \\
    $~S_{\max}~$        & Maximum shear rate normalised by $D_R^*$ &   $~0 - 30~$  \\
  \end{tabular}
  \caption{Dimensionless parameters in the present study. %{\color{red} YH: PARAMETERS NEED DO BE DOUBLE CHECKED}
  }
  \label{tab1}
  \end{center}
\end{table}

The depth of the channel in the present study is chosen to be $d(\equiv 2h)=0.5cm$, which is comparable to that used in \cite{durham2009disruption} ($d=1cm$).% and identical to the reference case found in \cite{ref2hwang2014bioconvection}. 
The range of the centreline velocity $U_c$ tested in the present study is $U_c=0-0.25cm/s$,
%${\color{red} 0-316 dimensionless. CORRECT THEDIMENSIONS}$,
leading to {$Re=0-6.28$}. The onset of gyrotactic trapping is strongly linked to the base flow vorticity at which an isolated cell will begin to tumble. For a given dimensionless gyrotactic time scale $\lambda$, such a vorticity can be calculated either in terms of the gyrotactic time scale $B$ or in terms of the rotational diffusivity $D_R^*$.
The critical spanwise vorticity (made dimensionless by the rotational diffusivity) at which a cell undergoing no random motion begins to tumble is $S_{crit}=4.4$, % \cite[for the value of $S_{crit}$, see][]{ref2hwang2014bioconvection},
 where the spanwise vorticity of the base flow is defined as
\begin{equation}\label{eq:2.5}
S(y)=-\frac{Sc Re}{D_R^*}\frac{dU_0}{dy}.
\end{equation}

We note that (\ref{eq:2.5}) is useful in characterising the flow rate in relation to the cell dynamics. Using (\ref{eq:2.5}), the base flow rate can be characterised by the maximum spanwise vorticity,
\begin{equation}
S_{\max}\equiv\max_yS(y),
\end{equation}
for which $S(y)=S_{\max}(=2ScRe/D_R)$ is attained at the upper wall (i.e. $y=1$). 

All the model parameters for the individual cell in the present study (e.g. swimming speed, sedimentation speed, cell volume, etc) are those for \emph{Chlamydomonas nivalis}.
%as in \cite{ref2hwang2014bioconvection}.
The range of the averaged cell number density considered in the present study is $N=1\times10^3-1\times 10^7 \mathrm{cells}/\mathrm{cm}^{3}$, which well includes $1.1\times 10^6\mathrm{cells}/\mathrm{cm}^{3}$ in the experiment of \cite{durham2009disruption}. Given the flow geometry and the parameters for the cell, this results in $Ra=10^{0} -10^{4}$. All the dimensionless parameters examined in the present study are summarised in table \ref{tab1}. %{\color{red}YH:PLEASE ALL THE STATEMENTS HERE VERY CAREFULLY (ESPECIALLY THE VALUES OF THE PARAMETERS)}

\section{Results and discussion}
\subsection{Basic state\label{sec:basicstate}}

\begin{figure*}[t!]
\vspace{-4mm}
    \centering
    \hspace{-10mm}
    \begin{subfigure}[H!]{0.4\textwidth}
        \centering        \includegraphics[width=1.1\textwidth]{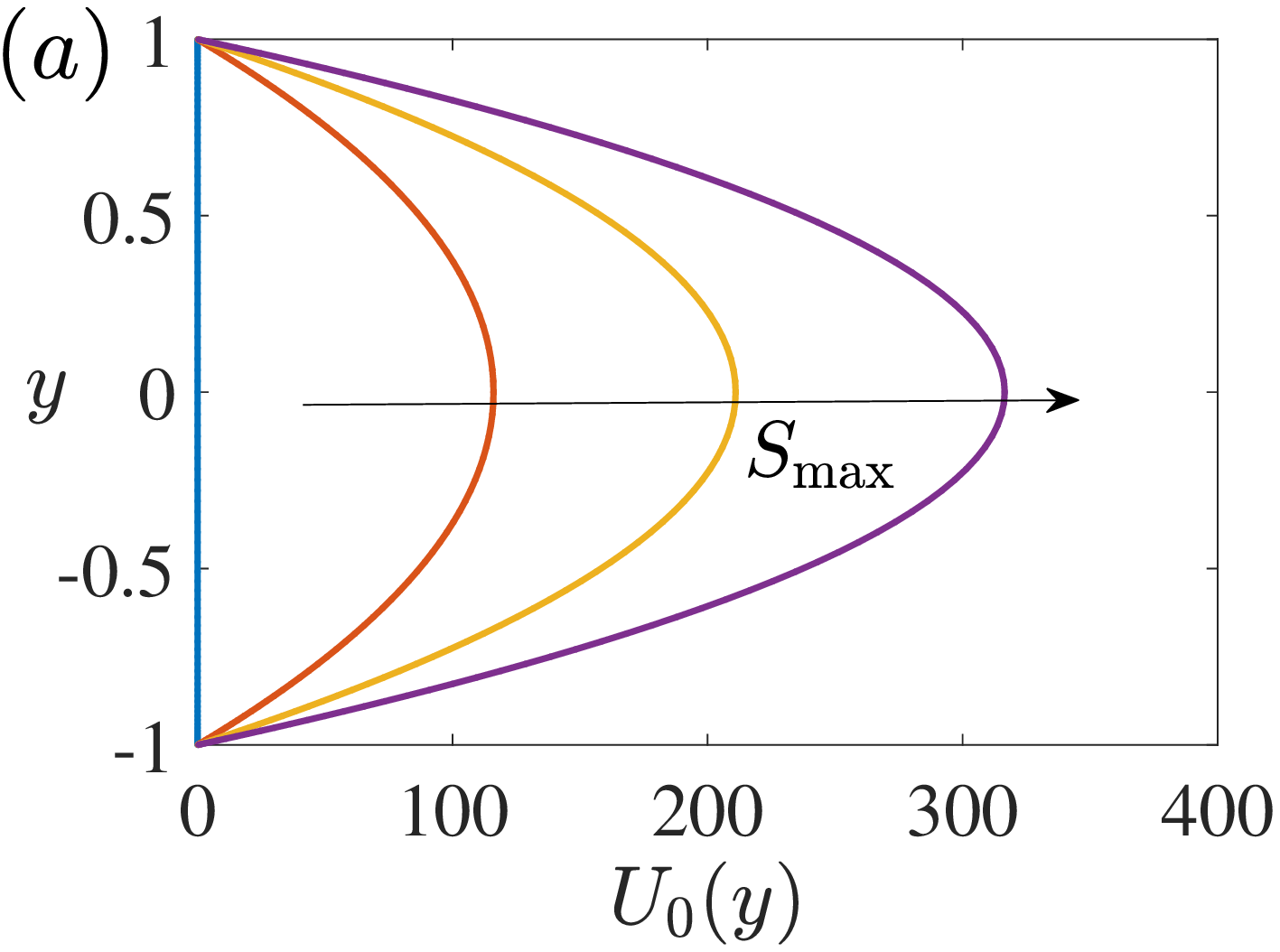}
        \label{Basicplota}
    \end{subfigure}%
     ~ \hspace{10mm}
    \begin{subfigure}[H!]{0.4\textwidth}
        \centering
        \includegraphics[width=1.1\textwidth]{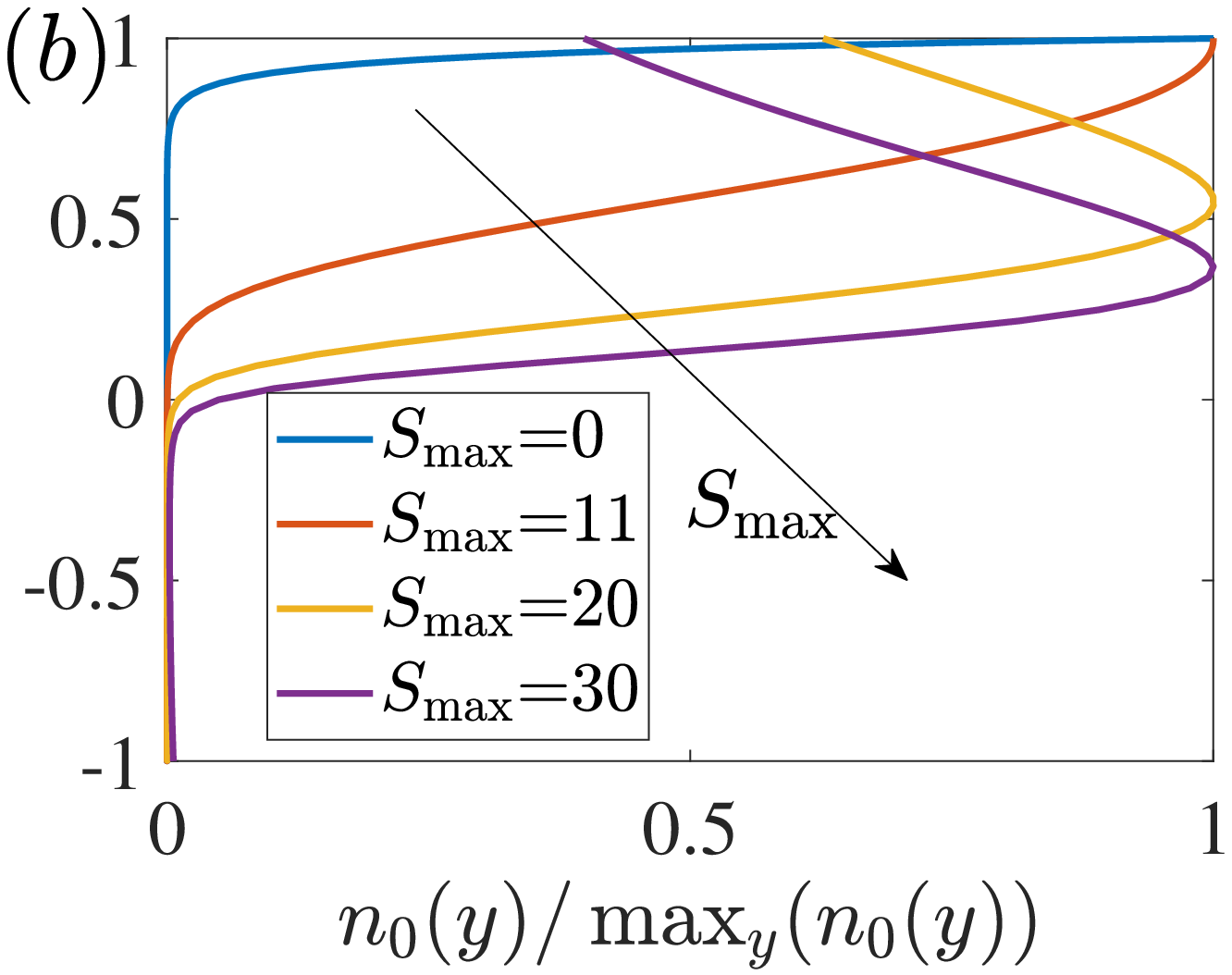}
        \label{Basicplotb}%\caption{}
    \end{subfigure}\\
    \caption{Profiles of the basic state with $S_{\max}=0, 11, 20, 30$: $(a)$ base flow; $(b)$ cell number density normalised by its maximum \label{basicplot}.}
\end{figure*}

The base flow and the corresponding cell number density distribution are plotted in figures \ref{basicplot}$(a)$ and $(b)$, respectively. As $S_{\max}$ is increased, the centreline velocity simply increases while maintaining the parabolic profile. In contrast, the profile of the cell number density experiences non-trivial changes with the increase of $S_{\max}$. To explain this feature, we start by making a few observations. Firstly, for a excessively large spanwise vorticity, the average upswimming speed of the cells should become smaller than the sedimentation speed, because the upswimming speed $V_c\langle e_2\rangle_0$ approaches zero in the limit of infinitely large vorticity \cite[see figure 3$a$ in][]{ref2hwang2014bioconvection}. In other words, $V_c\langle e_2\rangle_0\leq V_s$ for $S(y)\geq S_{s}$, where $S_s$ is the spanwise vorticity at which $V_c\langle e_2\rangle_0=V_s$. Note that, for the given modelling parameters, $S_{s}\simeq 11$ \cite[figure 7 in][]{ref2hwang2014bioconvection}. Secondly, the form of $n_0(y)$ in (\ref{basefloweqn2}) indicates that the sign of $V_c\langle e_2\rangle_0-V_s$ should be identical to that of $\mathrm{d}n_0/\mathrm{d}y$. It follows that the peak location of $n_0(y)$ is identical to the location where $S(y)=S_{s}$ (or equivalently $V_c\langle e_2\rangle_0=V_s$), suggesting the crucial role of the sedimentation speed in the gyrotactic trapping even if it is significantly smaller than the cell swimming speed \cite[note that $V_s\simeq 0.1V_c$; see][]{pedley2010instability}. 

Keeping these observations in mind, let us now observe $n_0(y)$ in figure \ref{basicplot}$(b)$ again. For $S_{\max}=0$, $n_0(y)$ is simply an exponentially growing function in the vertical direction because $\langle e_2\rangle_0$ and $D_{T0}^{22}$ in (\ref{basefloweqn2}) are constant. When $S_{\max}=11$, $S(y=1)\simeq S_s$. Therefore, $\mathrm{d}n_0/\mathrm{d}y \simeq 0$ at $y=1$. If $S_{\max}$ is increased further ($S_{\max}=20$), the peak location of $n_0(y)$, at which $S(y)=S_s$, now emerges in the region of $0<y<1$, exhibiting an equilibrium layer of the cells formed by the gyrotactic trapping. With a further increase of $S_{\max}$, the peak location of $n_0(y)$ is shifted further downward, as is shown for $S_{\max}=30$. 
 
\subsection{Linear stability analysis}
\begin{figure*}[h!]
 %\vspace{-8mm}
    \centering
    \hspace{-20mm}
    \begin{subfigure}[H]{0.4\textwidth}
        \centering        \includegraphics[width=1\textwidth]{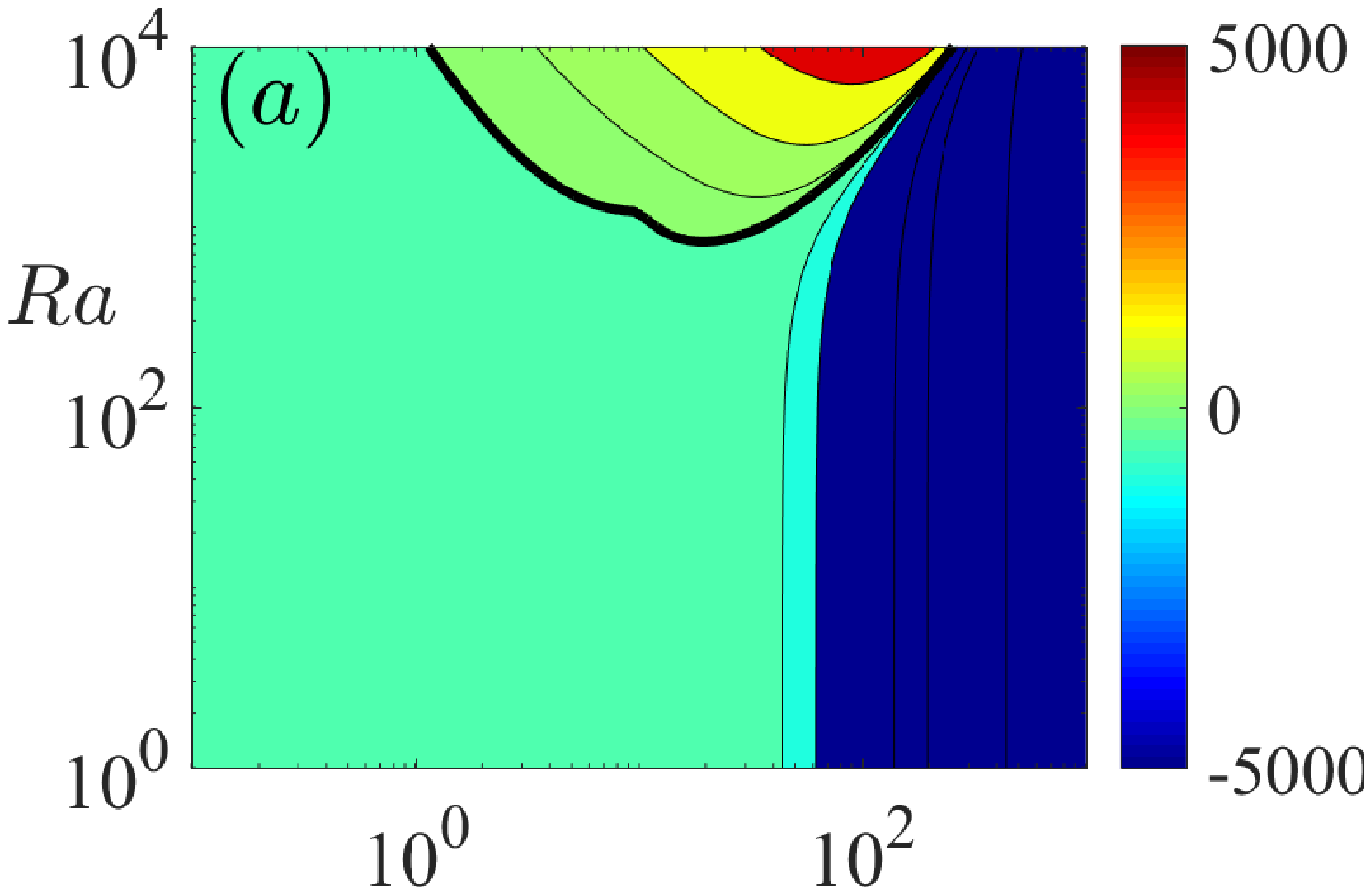}
 %       \caption{ $S=0$ for $\alpha\neq0$ and $\beta=0$ \label{ContPoiseuilleaS0}}
    \end{subfigure}%
     ~ \hspace{10mm}
    \begin{subfigure}[H]{0.4\textwidth}
        \centering
      \includegraphics[width=1\textwidth]{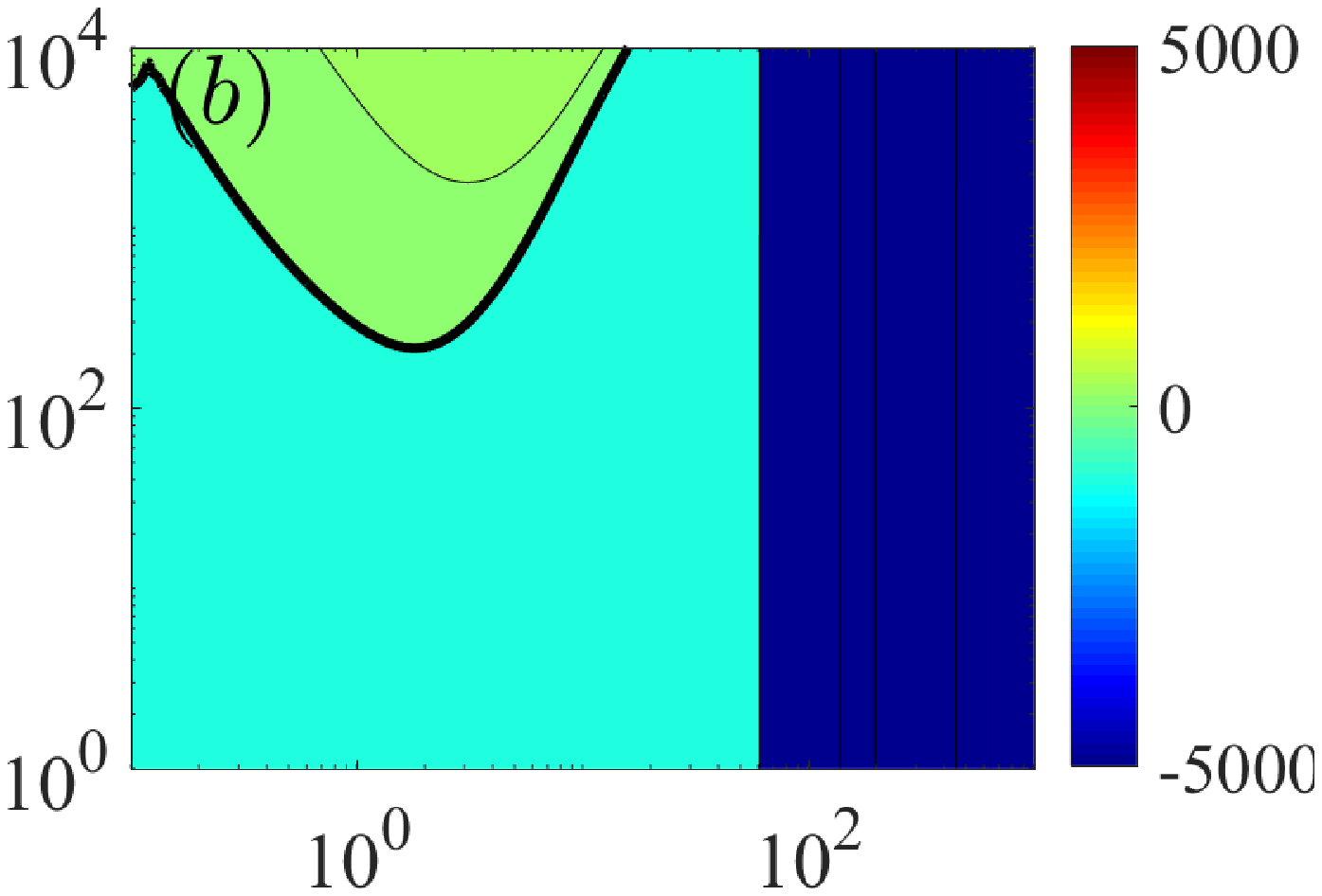}
  %      \caption{ $S=4$ for $\alpha\neq0$ and $\beta=0$\label{ContPoiseuilleaS4}}
    \end{subfigure}\\
        \vspace{-1mm}
    \hspace{-13mm}
    \begin{subfigure}[H]{0.4\textwidth}
        \centering        \includegraphics[width=1\textwidth]{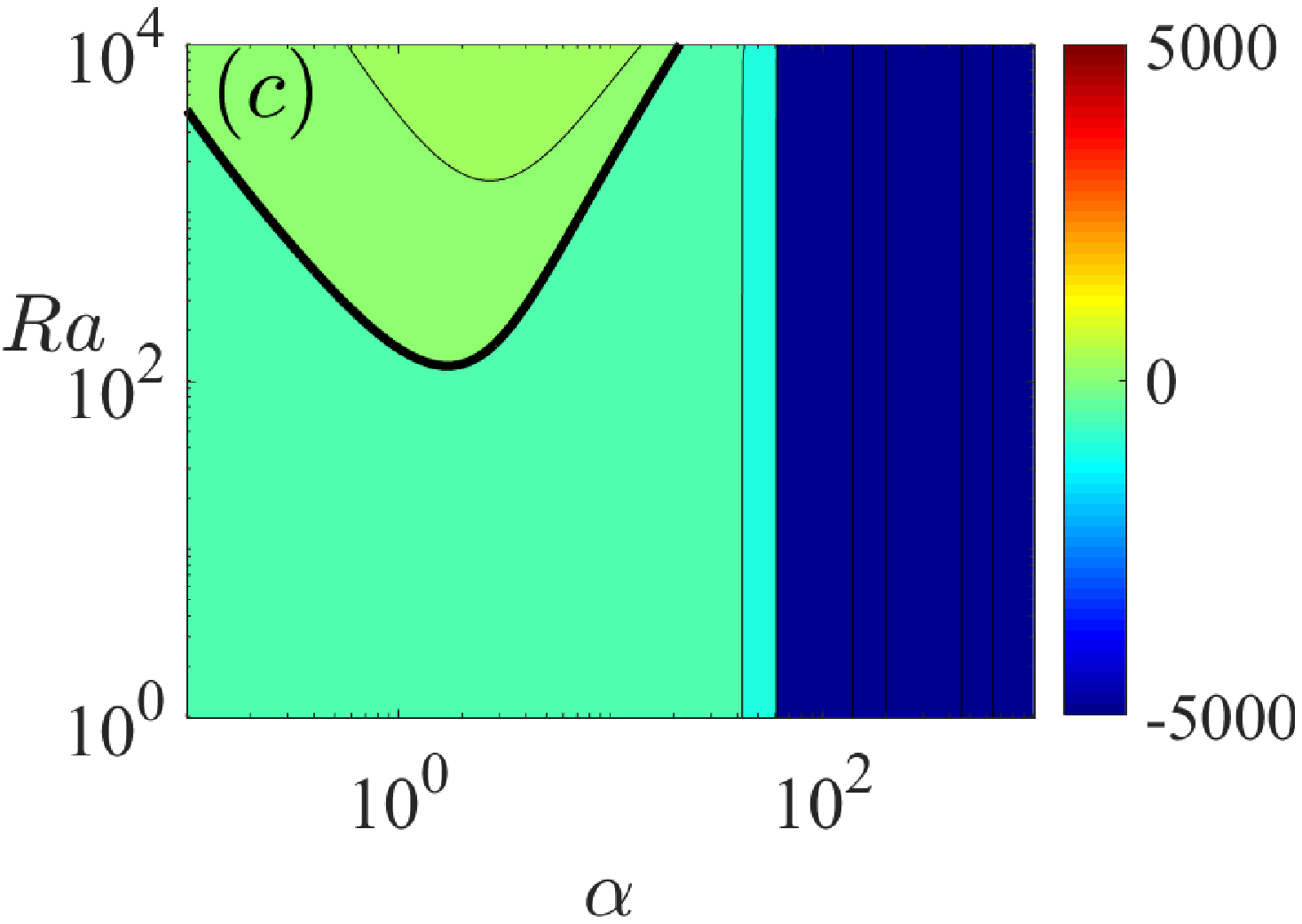}
   %     \caption{ $S=11$ for $\alpha\neq0$ and $\beta=0$\label{ ContPoiseuilleaS11}}
    \end{subfigure}
     ~ \hspace{9mm}
    \begin{subfigure}[H]{0.4\textwidth}
        \centering
      \includegraphics[width=1\textwidth]{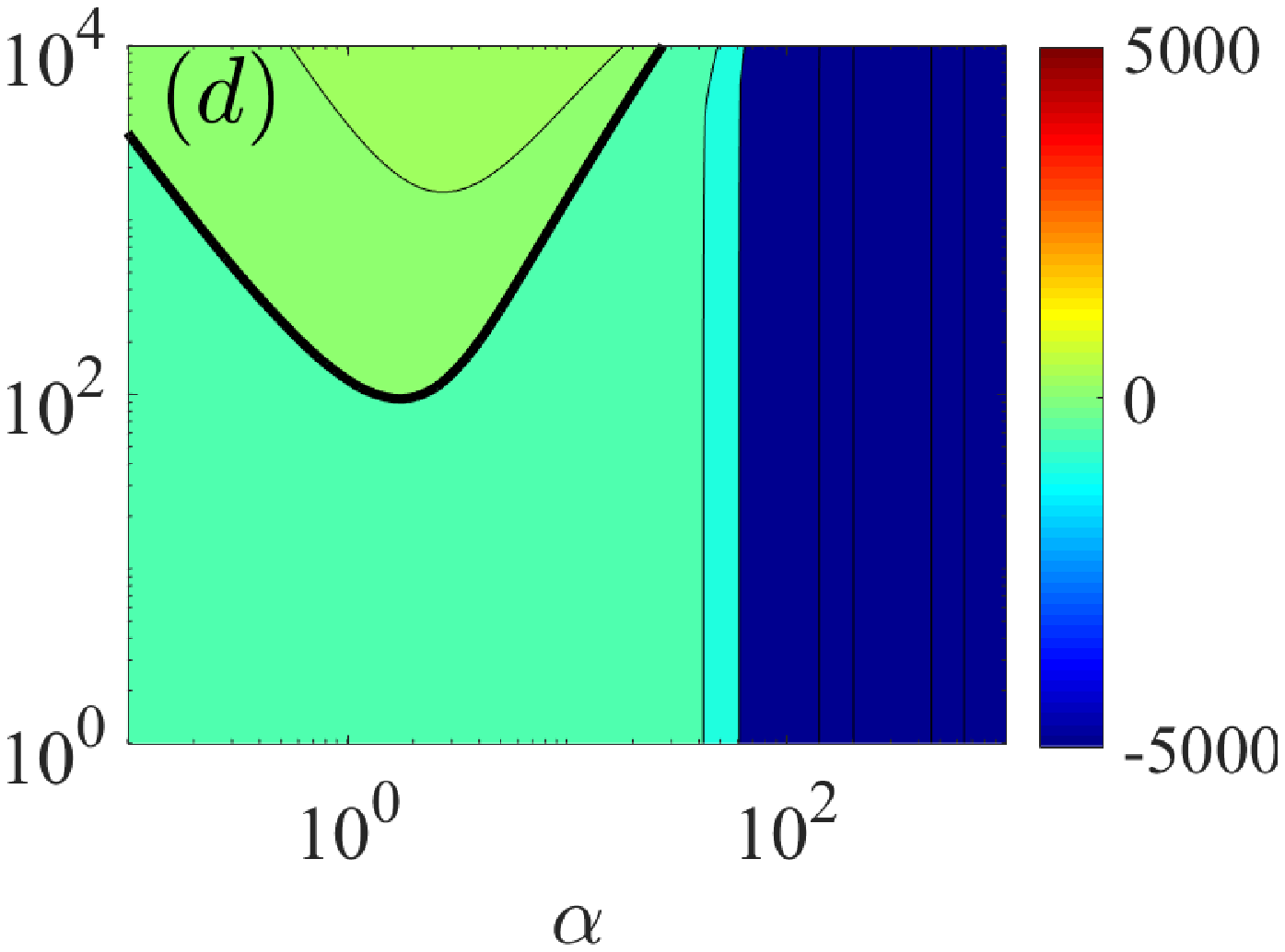}
    %    \caption{ $S=20$ for $\alpha\neq0$ and $\beta=0$\label{ContPoiseuilleaS20}}
    \end{subfigure}
            \vspace{-2mm}       
    \caption{Contours of the growth rate $\omega_i$ of the most unstable mode in the $Ra-\alpha$ plane for $\beta=0$: $(a)$ $S_{\max}=0$; $(b)$ $S_{\max}=11$; $(c)$ $S_{\max}=20$; $(d)$ $S_{\max}=30$. %YH: THE FONT SIZE IN THE FIGURE SHOULD BE INCREASED AND THEIR STYLE MUST BE IDENTICAL TO THOSE IN EQUATIONS.
\label{contourPoiseuille}}
        %    \vspace{-5mm}
\end{figure*}

\begin{figure*}[h!]
 \vspace{1mm}
 \label{Poiseuillealpha} 
       \centering
    \hspace{-20mm}
    \begin{subfigure}[H]{0.4\textwidth}
        \centering
  \includegraphics[width=1\textwidth]{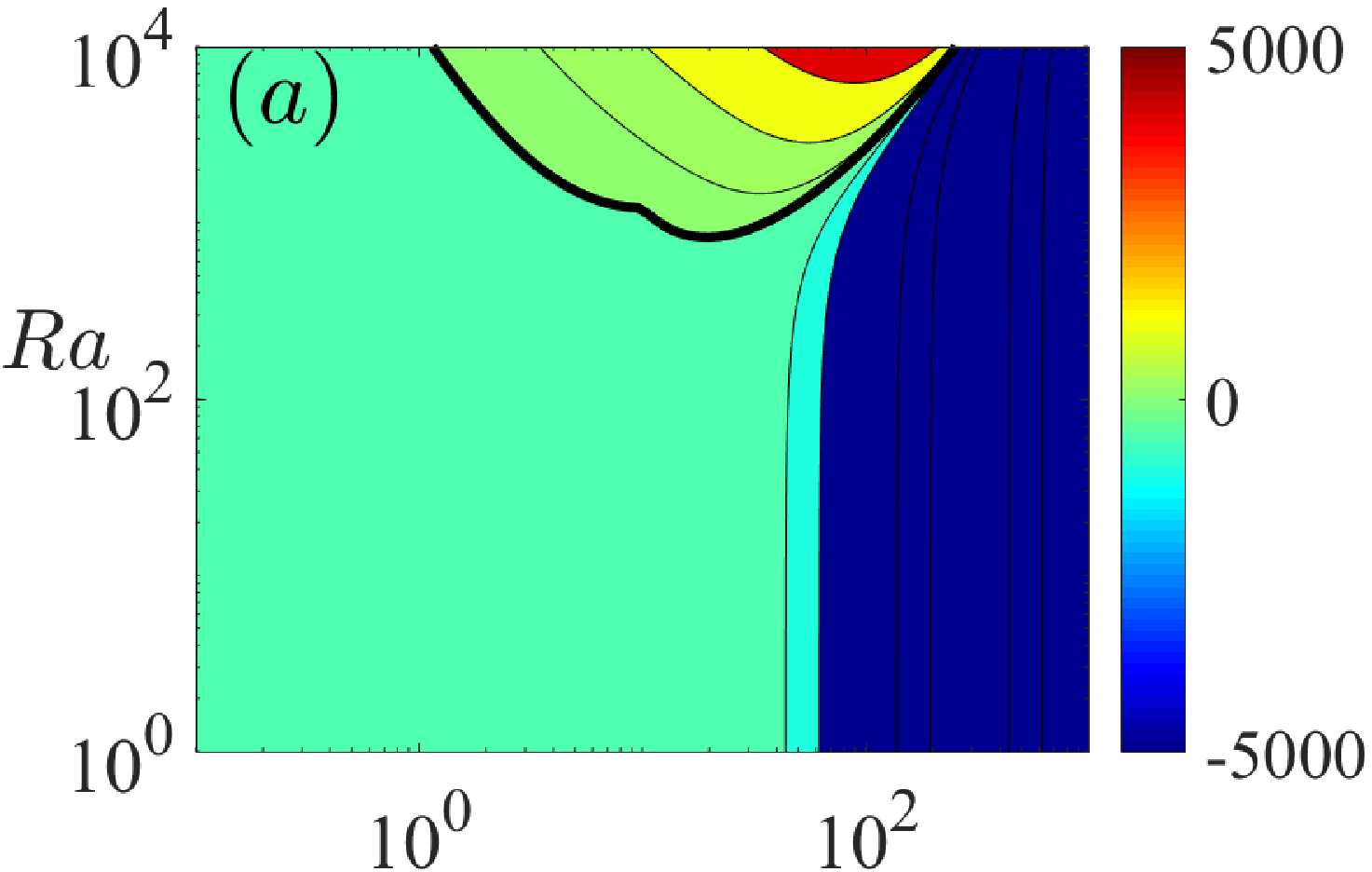}
%        \caption{ $S=0$ for $\alpha=0$ and $\beta\neq0$\label{ ContPoiseuillebS0}}
    \end{subfigure}%
     ~ \hspace{10mm}
    \begin{subfigure}[H]{0.4\textwidth}
        \centering
        \includegraphics[width=1\textwidth]{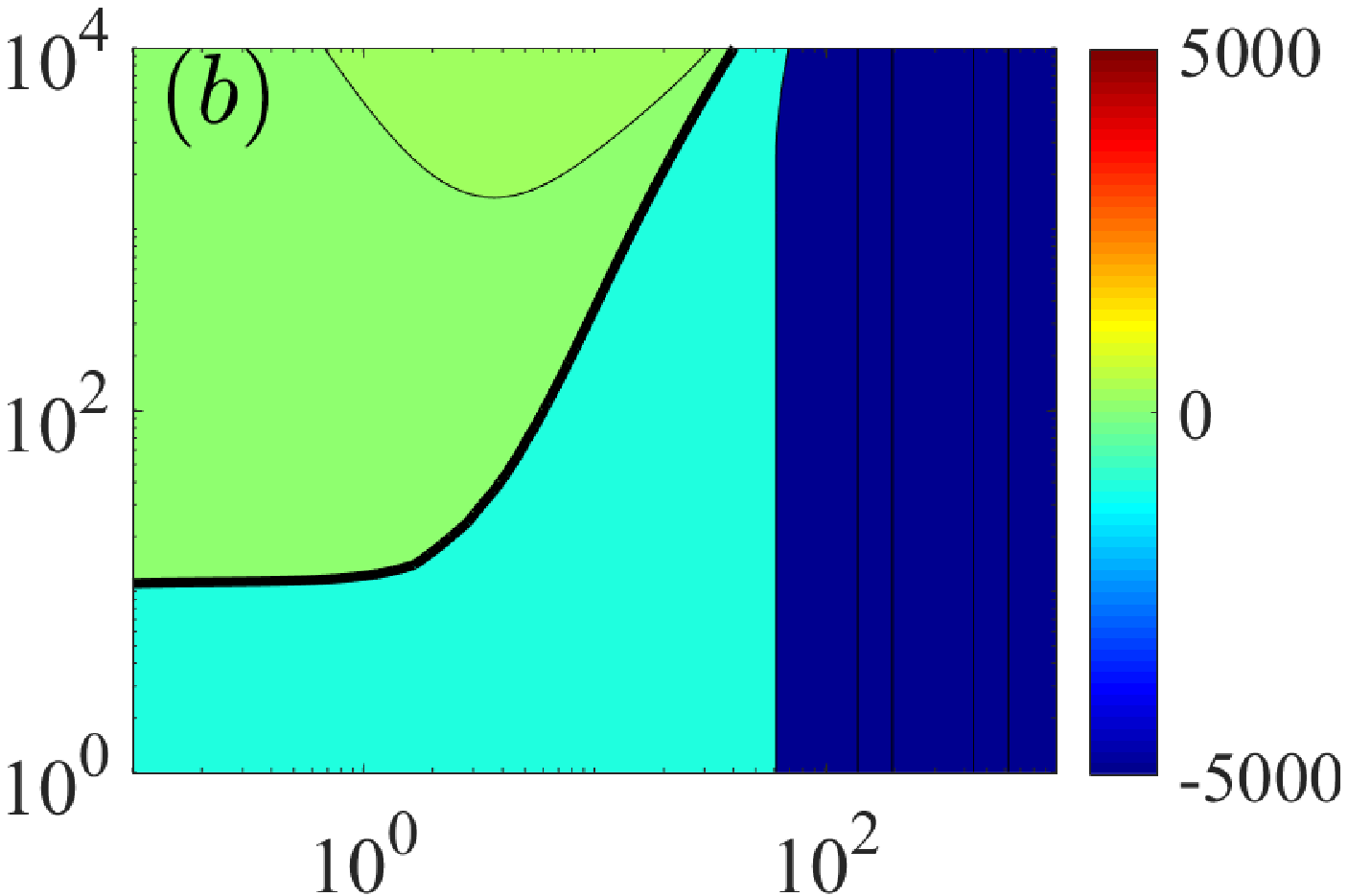}
 %       \caption{ $S=4$ for $\alpha=0$ and $\beta\neq0$\label{ ContPoiseuillebS4}}
    \end{subfigure}\\
            \vspace{-1mm}
       \centering
    \hspace{-13mm}
    \begin{subfigure}[H]{0.4\textwidth}
        \centering
  \includegraphics[width=1\textwidth]{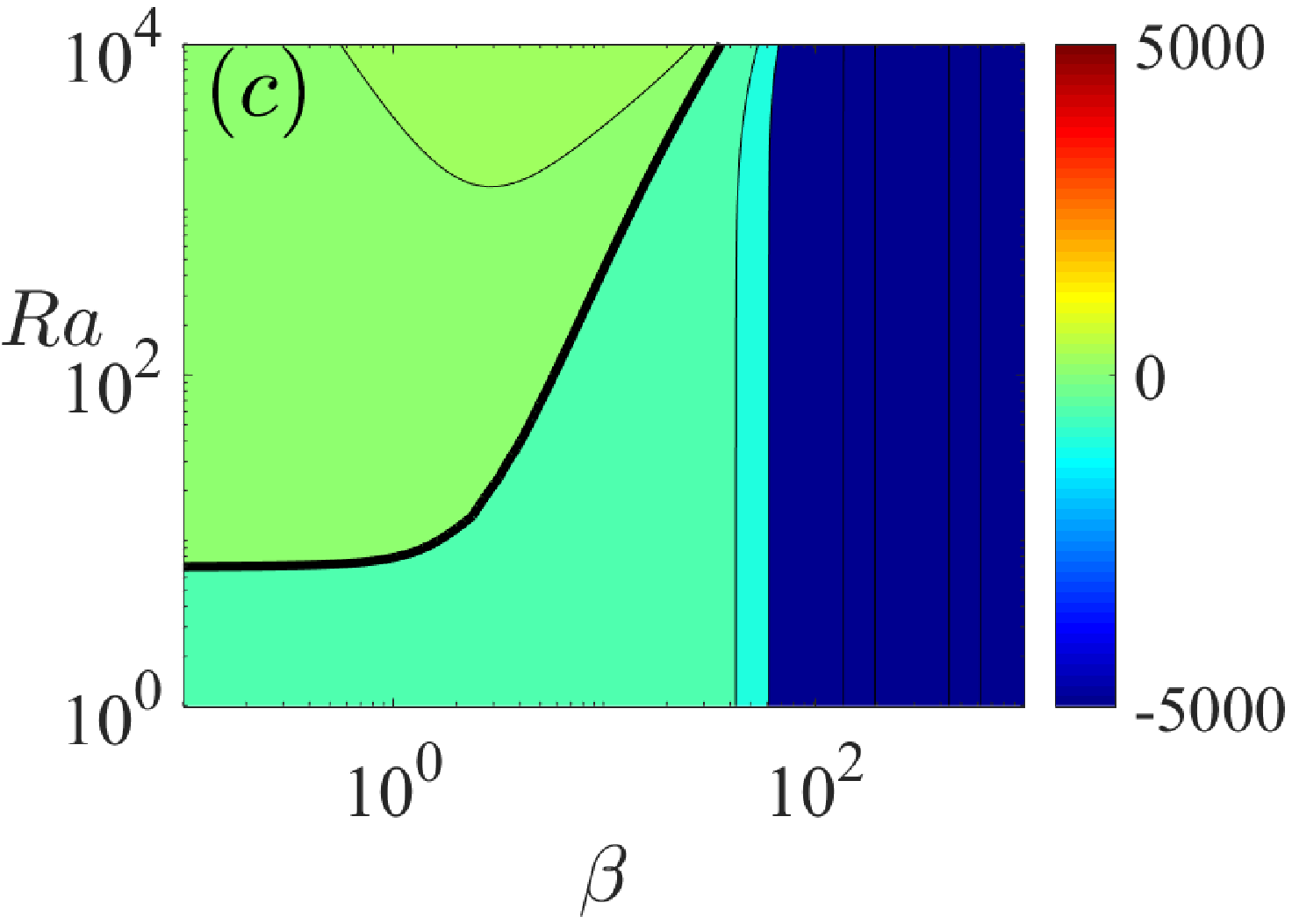}
  %      \caption{ $S=11$ for $\alpha=0$ and $\beta\neq0$\label{ ContPoiseuillebS11}}
    \end{subfigure}
      ~ \hspace{9mm}
    \begin{subfigure}[H]{0.4\textwidth}
        \centering
        \includegraphics[width=1\textwidth]{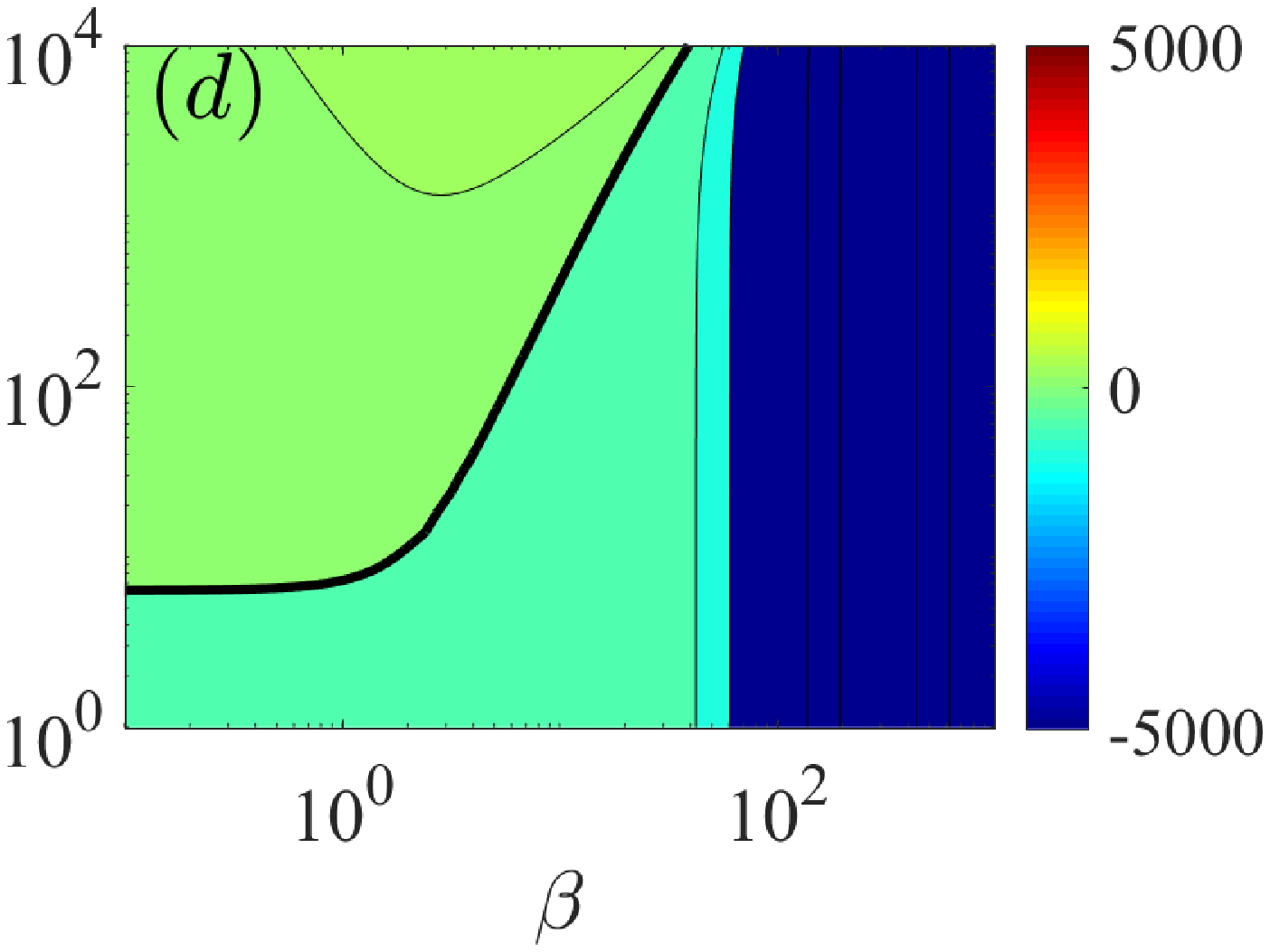}
   %     \caption{  $S=20$ for $\alpha=0$ and $\beta\neq0$\label{ContPoiseuillebS20}}
    \end{subfigure}\\
            \vspace{-2mm}
        \caption{Contours of the growth rate $\omega_i$ of the most unstable mode in the $Ra-\beta$ plane for $\alpha=0$: $(a)$ $S_{\max}=0$; $(b)$ $S_{\max}=11$; $(c)$ $S_{\max}=20$; $(d)$ $S_{\max}=30$. %YH: THE FONT SIZE IN THE FIGURE SHOULD BE INCREASED AND THEIR STYLE MUST BE IDENTICAL TO THOSE IN EQUATIONS. ALSO THE CONTOUR LINES IN FIGURE 3 ARE NOT IDENTICAL TO THOSE IN FIGURE 2. PLEASE CHECK.
\label{contourPoiseuillebeta}}
\end{figure*}

A linear stability analysis of the basic states in \S\ref{sec:basicstate} is now performed. The contours of the growth rate $\omega_i$ of the most unstable spanwise uniform ($\beta=0$) mode in the $Ra-\alpha$ plane are shown in figure \ref{contourPoiseuille} for  $S_{\max}=0,  11, 20  $ and $30$. For $S_{\max}=0$ (figure \ref{contourPoiseuille}$a$), the neutral stability curve and the contour plot are identical to those for stationary bioconvection in \cite{bees1998linear} and \cite{ref2hwang2014bioconvection}. On increasing $S_{\max}$, the instabilities at high streamwise wavenumbers ($\alpha>20$) are significantly damped, while the low wavenumber region ($\alpha<20$) is destabilised (figures \ref{contourPoiseuille}$b$-$d$). This tendency observed with increasing base-flow vorticity is fairly similar to that in the uniform shear flow \citep{ref2hwang2014bioconvection}. However, it should be mentioned in that case, even the low wavenumber region was also found to be completely stabilised once $S_{\max}>S_s$, whereas, in the present study, the region remains unstable even when $S_{\max}$ is roughly three times greater than $S_s$. Qualitatively, the same feature appears for the streamwise uniform instability mode ($\alpha=0$), as shown in figure \ref{contourPoiseuillebeta}. The only difference between this case and the spanwise uniform case is that the streamwise uniform mode exhibits higher growth rates at low spanwise wavenumbers ($\beta<20$) as $S_{\max}$ increases.

%\FloatBarrier

 %, while in figure  \ref{ ContPoiseuillebS8} the instability region is wider. 
 %Finally, the instability region for $S>11$ in Poiseuille flow is even larger than for $S=8$ (see figures \ref{ContPoiseuillebS15}-\ref{ContPoiseuillebS30}). 
%%%%%%%%%%%%%%%%%%%55

 %%%%%%%%%%%%%%
 \begin{figure*}[h!]
%%%%%%%  !!!!!!!! set(gcf, 'Renderer', 'opengl')!!!!!!
    \centering
    \hspace{-15mm}
    \begin{subfigure}[H]{0.8\textwidth}
        \centering  \includegraphics[width=1.1\textwidth]{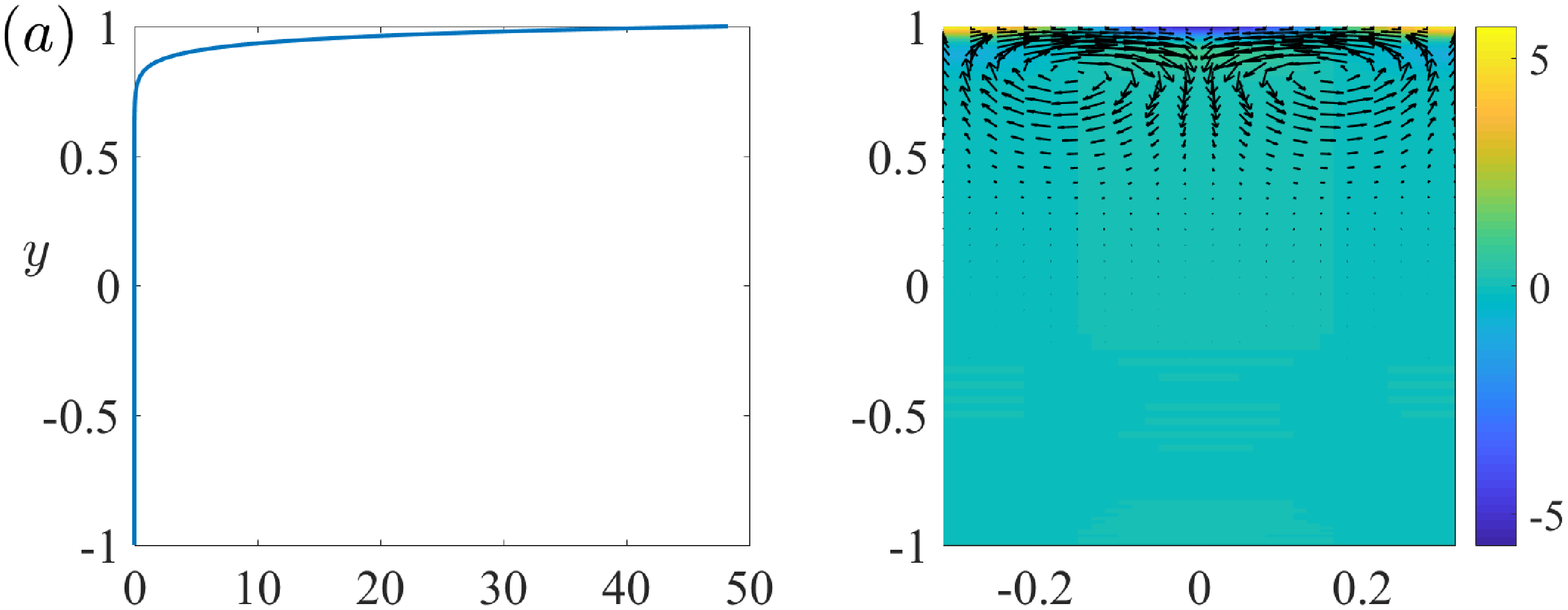}
     \label{ Poiseuillebeta0}  %  \caption{ $S=0$.\label{Poiseuillebeta0}}     
    \end{subfigure}
\\
\vspace{-1mm}
       \centering
    \hspace{-10mm}
    \begin{subfigure}[H]{0.8\textwidth}
        \centering   \includegraphics[width=1.1\textwidth]{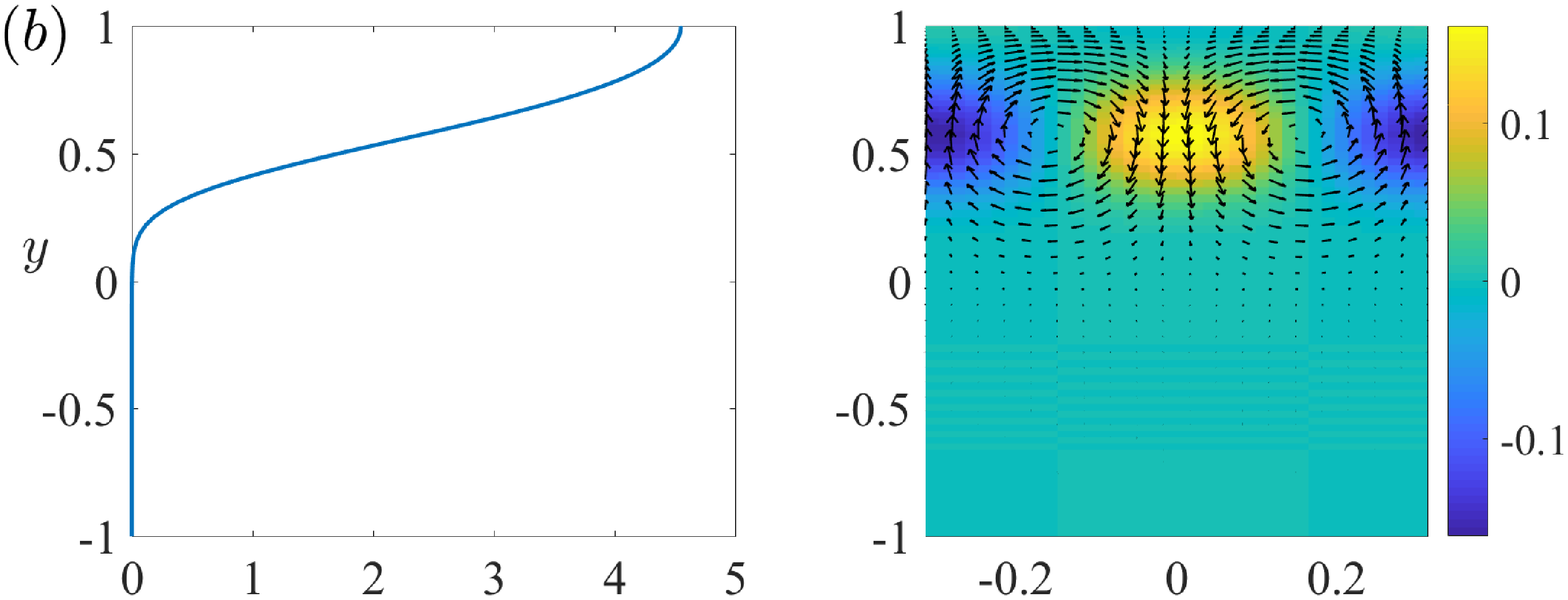}
     \label{ Poiseuillebeta11}  % \caption{ $S=4$.\label{ Poiseuillebeta4}}
    \end{subfigure}%
\\
\vspace{-1mm}
       \centering
    \hspace{-10mm}
    \begin{subfigure}[H]{0.8\textwidth}
        \centering   \includegraphics[width=1.1\textwidth]{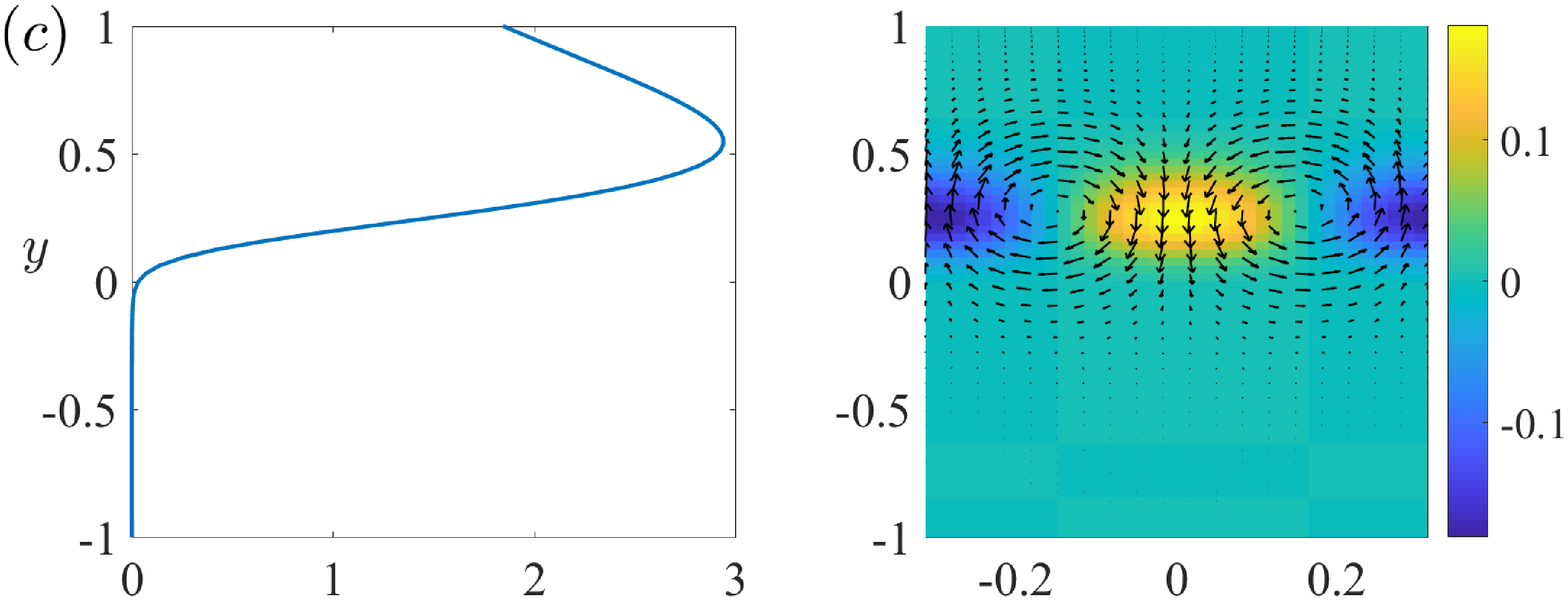}
     \label{ Poiseuillebeta20}   %\caption{  $S=11$.\label{ Poiseuillebeta8}}
    \end{subfigure}%
    \\
\vspace{-1mm}
       \centering
    \hspace{-10mm}
    \begin{subfigure}[H]{0.8\textwidth}
        \centering
  \includegraphics[width=1.1\textwidth]{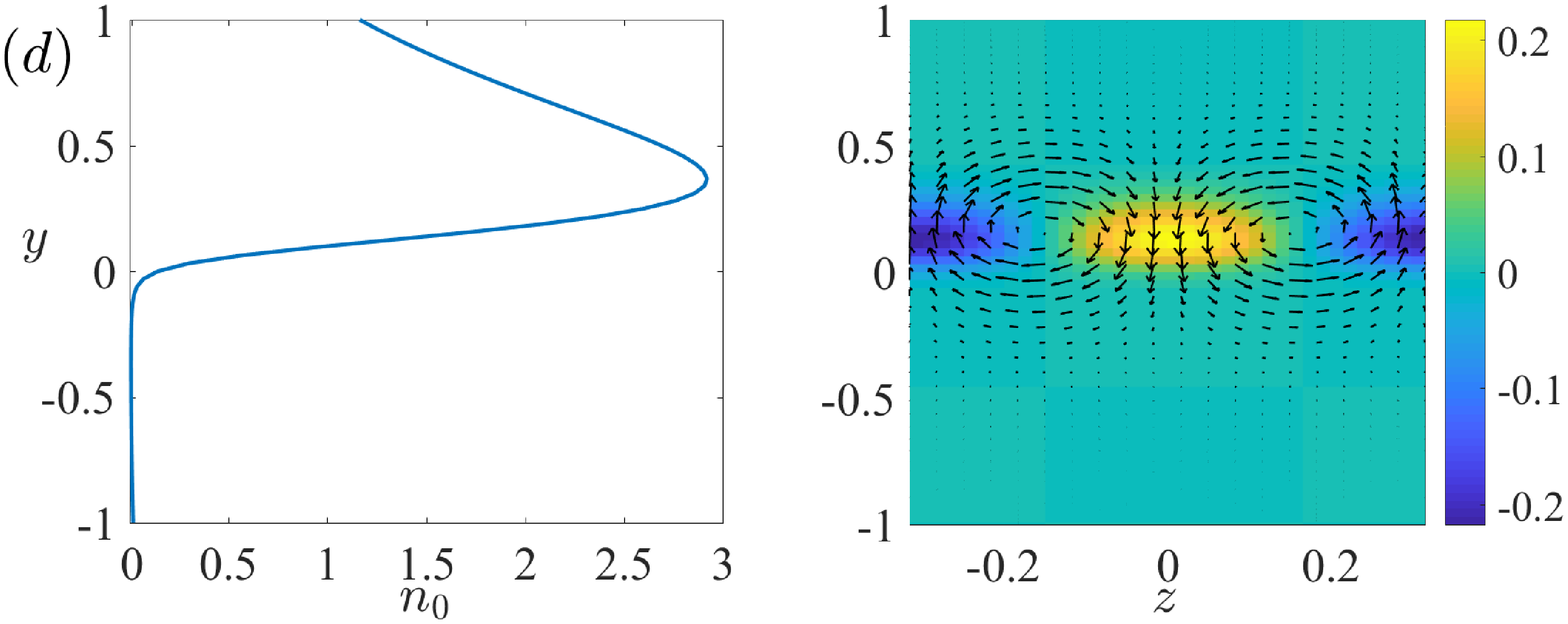}
       \label{ Poiseuillebeta30} %\caption{  $S=20$.\label{ Poiseuillebeta20}}
    \end{subfigure}%
    \vspace{-1mm}\label{Poiseuillebeta}
    \caption{Basic-state cell number density (left column) and cross-streamwise view of the corresponding eigenfunction for $\alpha=0$, $\beta=10$ and $Ra=2000$ (right column): $(a)$ $S_{\max}=0$; $(b)$ $S_{\max}=11$; $(c)$ $S_{\max}=20$; $(d)$ $S_{\max}=30$. Here, the velocity vectors show the spanwise and wall-normal velocity perturbation, and the contours indicate the perturbed cell-number density. The velocity perturbation and the perturbed cell-number density are normalised by -max$|\hat{v}|\big|_{z=0}$. \label{PoiseuillebetaEigfun} %YH: THE FONT SIZE IN THE FIGURE SHOULD BE INCREASED AND THEIR STYLE MUST BE IDENTICAL TO THOSE IN EQUATIONS. 
    }
\end{figure*}
%%%%%%%%%%%%%%%%%%%%%%%%%%%%
%%%%%%%%%%%%% 
%\section{Discussion\label{sec:discussion}}
%Our study has focused on  how parabolic shear affects the linear stability of shallow layers of gyrotactic microorganism suspensions. We found that high wave-number instabilities are significantly damped when increasing shear, and that shear  acts increasingly destabilising for low-wavenumbers. We further found that instability regions do not disappear with increased shear. A further detail to note, is that with a streamwise uniform mode, the instability region increases for $S_{max}\geq11$. Based on our observations, it is for  $S_{max}>11$ onwards that the maximum basic state cell number density is no longer at the upper wall of the channel i.e. the cell layer forms below the top wall. 
%In fact, as $S_{max}$ increases , the swimmers are forced to cluster and form a layer.
%The existence of sheared rotating rolls below  prevents the layer from sinking more, essentially trapping the swimmers from below, while the high shear from above traps them from above. Increasing shear further only leads to the compression of the corotating rolls as the shear required for swimmers to tumble is reached at a lower vertical position. Note also, that increasing shear does not lead to the disintegration of the corotating rolls, as seen in figure \ref{PoiseuillebetaEigfun}.

%%%%%%%
\subsection{Physical mechanism of the instability}

To understand the origin of the persistent instability,  even at a considerably large base-flow vorticity, we first explore how the basic-state cell number density profile is correlated with that of the eigenfunction of the instability mode. Figure
\ref{PoiseuillebetaEigfun} shows the basic-state cell number density (left) and the cross-streamwise structure of the most unstable eigenmode (right) for $\alpha=0$, $\beta=10$ and $Ra=2000$. As explained previously, the maximum $n_0$, which indicates the position of the cell layer formed by the gyrotactic trapping, shifts downwards with increasing $S_{\max}$. Interestingly, the vertical location, in which the eigenmode appears in the form of counter-rotating rolls with the corresponding cell-number density field, also moves downwards together with the peak location of $n_0$. Furthermore, for all $S_{\max}$ considered here, the eigenmode of the instability consistently emerges in the region where the basic cell-number density is unstably stratified (i.e. $dn_0/dy>0$), suggesting that the instability would be associated with the gravitational overturning mechanism observed in Rayleigh-B\'enard convection and the Rayleigh-Taylor instability. 

 \begin{figure*}[h!]
    \centering
       % \vspace{-7mm}
    \hspace{-20mm}
\begin{subfigure}[H]{0.4\textwidth}
        \centering
        \includegraphics[width=1.1\textwidth]{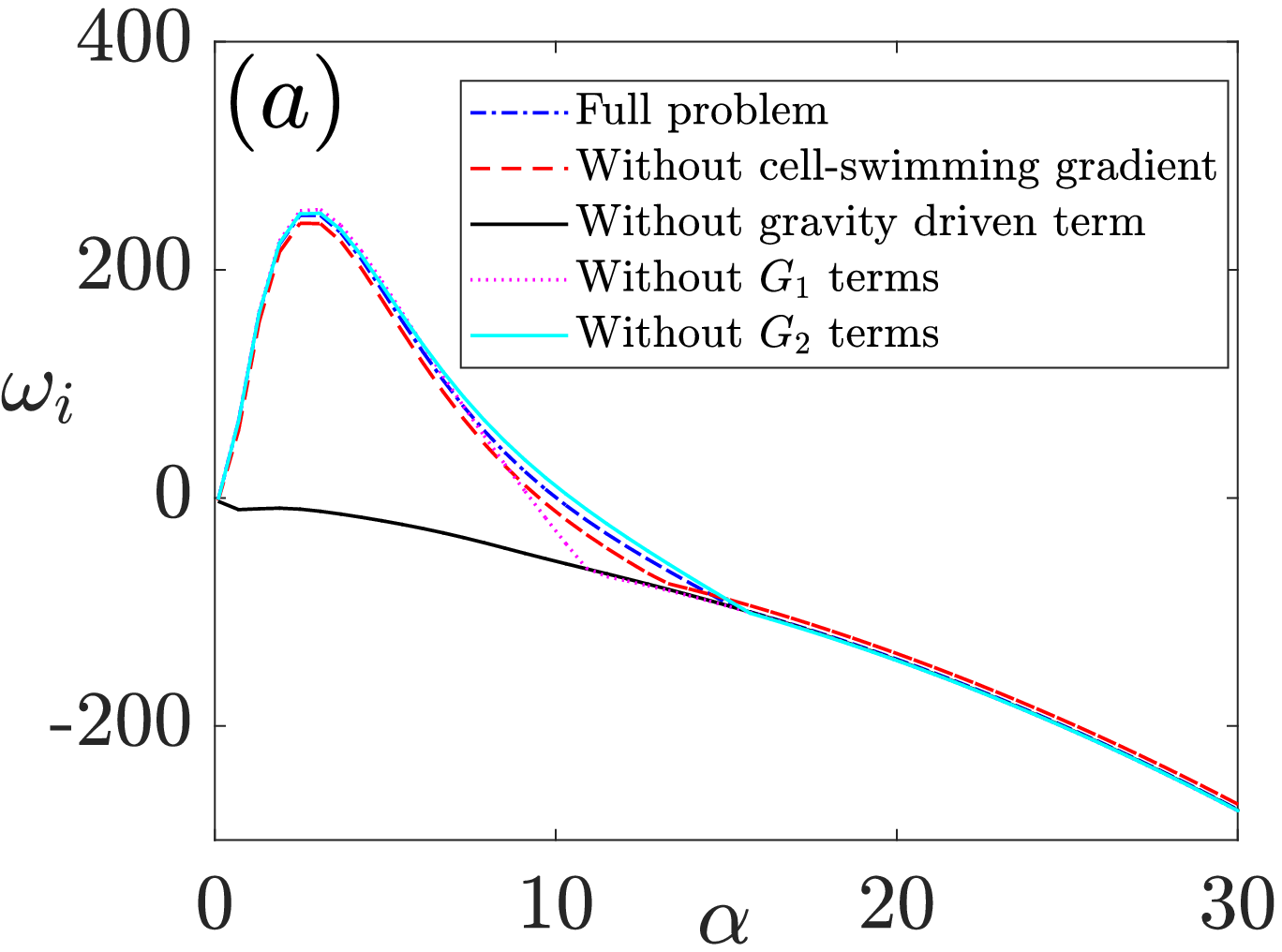}
     %   \caption{$\alpha$ vs. $\omega_i$\label{ SuppressContPoiseuilleaS20}}
    \end{subfigure}
    ~ \hspace{4mm}
    \begin{subfigure}[H]{0.4\textwidth}
        \centering
        \includegraphics[width=1.1\textwidth]{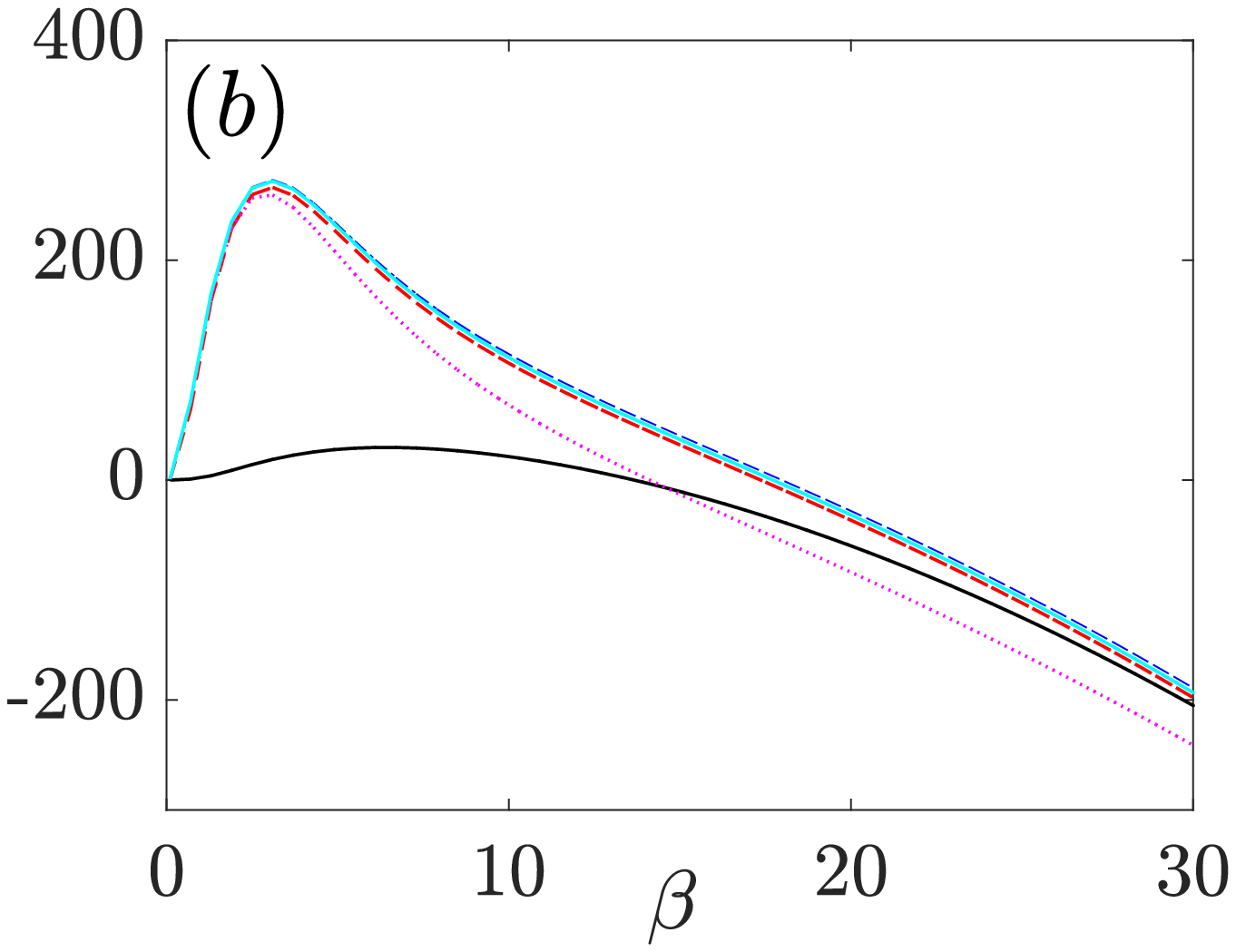}
      %  \caption{$\beta$ vs. $\omega_i$ \label{ SuppressContPoiseuillebS20}}
    \end{subfigure}
    \caption{Examination of the instability mechanisms for $S_{\max}=20$ and $\Ra=2000$: $(a)$ $\omega_i$ vs $\alpha$ with $\beta=0$; $(b)$ $\omega_i$ vs $\beta$ with $\alpha=0$. %YH: PLEASE ALSO CHECK IF THE CAPTION IS CORRECT. %YH: WE NEED TO CHANGE THE RANGE OF WAVENUMBERS FROM 0 TO 30. WE DON'T NEED TO PRESENT THE INFLECTIONAL ONE HERE. 
\label{SuppresscontourPoiseuillealpha}}
\end{figure*}

The precise mechanism of the instability persisting at large $S_{\max}$ is further investigated by carefully examining each term of (\ref{lns}). %As in \cite{ref3hwang2014stability}, 
Four different physical mechanisms of instability are identified in the present case: 1) gravitational instability ($\mathcal{D}n_0$ in the third row of $(2.4a)$); 2) gyrotactic instability (all the terms with $G_1$ in $(2.4e)$ and $(2.4f)$); 3) diffusion-oriented instability (all the terms with $G_2$ in $(2.4e)$ and $(2.4f)$); 4) instability caused by the spatial gradient of cell-swimming vector field ($V_c\mathcal{D}\langle{e_2}_0\rangle_0$ in $(2.4d)$). The first three mechanisms here were previously shown to play almost equally important roles in bioconvection instability \citep{ref2hwang2014bioconvection}, while the last one was shown to be the central mechanism for the blip instability in downward channel flow \citep{ref3hwang2014stability}. Given the scope of the present study for the instability of gyrotactic trapping, here we focus on the instability emerging for $S_{\max}>S_s$. In such a case, the spanwise vorticity of the base flow would be fairly large in most of the vertical domain, and this yields all $\xi_i$'s in (\ref{lns}) fairly small \cite[the values of $\xi_i$ quickly diminish to zero as $S(y)\rightarrow \infty$; see figure 6 in][]{ref2hwang2014bioconvection}. We note that the $\xi_i$'s appear with $G_1$ and $G_2$ throughout (\ref{lns}), indicating that the gyrotactic and diffusion-oriented mechanisms are unlikely to be very active for $S_{\max}>S_s$.

The discussion given above now suggests that the potential instability mechanism of the layer formed by gyrotactic trapping would originate from the gravitational mechanism and/or the one by the spatial gradient of cell-swimming vector field. To check this, we perform a numerical experiment for $Ra=2000$ and $S_{\max}=20$, in which linear stability is examined by suppressing each of the terms discussed above individually. As shown in figure \ref{SuppresscontourPoiseuillealpha}, the dominant contribution of the instability is made by the term associated with the gravitational mechanism (i.e. $\mathcal{D}n_0$ in the third row of $(2.4a)$), as its suppression leads to complete stabilisation. However, the one from the spatial gradient of cell-swimming vector field is found to play no role because its suppression hardly changes the growth rate, suggesting that the dominant instability mechanism in the present study is the gravitational one.

\section{Concluding remarks}
Our study indicates that while a layer of bottom-heavy cells formed by gyrotactic trapping is indeed an equilibrium solution to the continuum equations describing such a suspension, this layer is also linearly unstable.  The high shear rate ($S_{\max}\geq 20$) critical Rayleigh number for this instability is two orders of magnitude lower than that for typical bioconvection.  This implies that the gyrotactically-trapped layer would be unstable at fairly low cell concentrations. In a suspension for the depth $d(=2h)=0.5\textrm{cm}$, bioconvection occurs at $N\simeq 10^6 \textrm{cells}/\textrm{cm}^{3}$. Based on our results, a gyrotactically-trapped layer would be unstable only at $N\simeq 10^4 \textrm{cells}/\textrm{cm}^{3}$ for the same depth. {It is interesting to note that this value of the cell concentration for the onset of the instability is fairly close to $N\simeq 10^3$--$ 10^4 \textrm{cells}/\textrm{cm}^{3}$ observed in thin layers at Monterey Bay \citep{jimenez1987relations, steinbuck2009observations}. As mentioned in the introduction, the thin layers of phytoplankton often develop in  regions where turbulence is  suppressed by strong density stratification \citep{dekshenieks2001temporal}. This suggests that the instability observed in the present study might be a mechanism that limits the cell concentration of thin layers in  aqueous environments where turbulence is  weak.} The physical mechanism of this instability is also robust -- as in bioconvection, it is a simple gravitational instability caused by the dense layer formed by gyrotactic trapping. {Ecologically, it is unclear why the cells would exhibit such collective behaviour. However, this might be an evolutionary outcome that the cells have developed to prevent the development of high concentrations. High cell concentrations could lead to a competitive environment for nutrient uptake, and could weaken cell swimming capabilities through hydrodynamic interactions between nearby cells, leaving the population susceptible to predators.}

{Despite this encouraging comparison, in the experiment of \cite{durham2009disruption}, the formation of gyrotactically-trapped cell layers was observed in a `time-averaged' sense even for $N\simeq 10^6\textrm{cells}/\textrm{cm}^{3}$ \cite[see figure 2$a$ in][]{durham2009disruption}.} In light of our current findings, this observation suggests the importance of the nonlinear evolution of the instability, especially as a function of the averaged cell number density. It is certainly possible that the instability is not strong enough at low average cell number densities (or Rayleigh numbers) to strongly disturb the layer formed by gyrotactic trapping. In other words, at low averaged cell number densities, the relatively weak instability may give rise to unsteady layer dynamics (as was also confirmed by private communication with W. M. Durham) with an overall time-averaged shape. Notwithstanding, there may be a more intricate interplay between cellular gyrotaxis and the evolution of the cell-layer instability that can only be ascertained through an exploration of the fully non-linear regime. {This can probably be studied with  classical weakly non-linear stability analysis as well as with full-nonlinear simulations of the continuum model. Performing more extensive and carefully controlled experiments on this issue would also be highly desirable.} 

{Finally, the typical thickness of the layer formed by gyrotactic trapping would be estimated by $l\sim D_v/V_c^*$. For \emph{C. nivalis}, $l \sim O(0.1-1\textrm{mm})$ \cite[][see also figure \ref{basicplot}$a$]{pedley2010instability} and the time scale of the instability would be greater than $O(1\textrm{s})$ for $N<10^6 \textrm{cells}/\textrm{cm}^{3}$ (figure \ref{SuppresscontourPoiseuillealpha}). We note that the typical Kolmogorov length and time scales in oceans are $\eta \sim O(0.1-10\textrm{mm})$ and $\tau_\eta \sim O(0.1-10^2\textrm{s})$ \citep{durham2013tur}. This implies that turbulent mixing in oceans would easily disrupt the layer formation process as well as its instability, consistent with the observation by \cite{dekshenieks2001temporal}. However, precisely to what extent and how this would happen needs to be understood. Other interesting avenues for further research would also include the role of hydrodynamic interactions as well as the effects of changes in the swimmers' physical properties (shape, distribution of mass, rigidity of the body). The hydrodynamic interactions between the cells can alter the precise shape of the layer as they would impact on diffusivity and rheology of the suspension \citep{ishikawa2007rheology}. The shape of the cell can also affect the dynamics of the suspension. For example, the layer may form at lower shear rates for elongated cells due to enhanced sedimentation by the vertical shear \citep{clifton2018enhanced}. The instability for such a layer also deserves a future investigation.}

\section*{Acknowledgement}
We gratefully acknowledge Dr W. M. Durham, who kindly provides a detailed explanation on his early experimental observation \citep*{durham2009disruption}. {We are grateful to the anonymous referees of this paper for their constructive comments on the original manuscript.} 

\bibliographystyle{jfm}
\bibliography{jfm-instructions}
\end{document}